\documentstyle[11pt,paspconf,epsf]{article}
\def\edcomment#1{\iffalse\marginpar{\raggedright\sl#1\/}\else\relax\fi}
\reversemarginpar

\begin{document}
\title{Time dependent models for Li, Be and B production in the early 
Galaxy\footnotemark}

\vspace{-15pt}
\author{E. Parizot and L. Drury}
\affil{Dublin Institute for Advanced Studies,
Dublin 2, Ireland}

\footnotetext{to appear in ``Galaxy Evolution: connecting the 
distant universe with the local fossile record'', eds.  M. Spite and 
F. Crifo, Les Rencontres de l'Observatoire de Paris, Kluwer, 
Dortrecht}

\vspace{-5pt}
\begin{abstract}
We calculate the light element production induced by the explosion of 
an isolated supernova in the ISM. We use a time-dependent model and 
consider energetic particles accelerated at the forward (process~1) 
and reverse (process~2) shocks.  Both processes are primary, but are 
shown to underproduce Be and B. The reasons for this failure are 
analyzed and used to propose a possible alternative, based on the 
acceleration of particles inside superbubbles.
\end{abstract}

\vspace{-20pt}
\section{Introduction}

Boosted by a wealth of new data on the Be and B abundances in very 
metal-poor stars of the Galactic halo (e.g.  Gilmore et al.  1992; 
Duncan et al.  1992,1997; Ryan et al.  1994; Edvardsson et al.  1994; 
Kiselman \& Carlsson 1996; Molaro et al.  1997; Garcia-L\'opez et al.  
1998), theoretical studies of the light element production and 
evolution in the early Galaxy have considerably developed during the 
past few years, and contributed to renew the nature of the questions 
asked in this context (e.g.  Feltzing \& Gustafsson 1994; Reeves 1994; 
Cass\'e et al.  1995; Fields et al.  1995; Ramaty et al.  1996,1997; 
Vangioni-Flam et al.  1998).  While everyone working in this field 
seems to agree that the Be atoms observed in the halo stars were 
synthesized by spallation reactions of C and O involving energetic 
particles accelerated in the interstellar medium (ISM), it is still 
not clear what kind of material is actually accelerated, where the 
acceleration takes place, by which mechanism, and how efficient it is.

As a guide towards the answer to such questions, special attention has 
been paid to the shape of the evolution of Be and B abundances in the 
early Galaxy, as a function of metallicity.  Indeed, contrary to what 
had been expected, these light element abundances show a linear growth 
with respect to Fe, indicating that Be and B are produced by a primary 
process.  By this, it is meant that Be and B production does not 
depend on the ISM metallicity, and is intrinsically (though probably 
indirectly) linked to the activity of massive stars, just as the C, O 
or Fe production, i.e.  independently of a prior enrichment of the 
ISM.

To account for these new data, several primary mechanisms have been 
proposed in which the light elements are produced through the 
spallation of energetic C and O nuclei accelerated out of freshly 
synthesized material (supernova ejecta, winds of Wolf-Rayet stars, 
etc.)  and interacting with the ambient (metal-poor) ISM. These 
mechanisms reproduce quite well the qualitative behavior of Be and B 
as compared to the common metallicity tracers (they are made for 
this).  However, the question of energetics, i.e.  quantitative 
agreement with the data, proves very difficult (e.g.  Ramaty et al.  
1997) and is still mostly unsolved.  Therefore, while concentrating on 
the shape of Be and B Galactic evolution admittedly gave important 
clues towards the identification of their origin, we feel that the 
problem of energetics should now be addressed more directly, as the 
constraint it provides proves powerful enough to rule out many 
qualitatively acceptable mechanisms.

In this contribution, we study two primary processes which could be 
expected to be especially efficient in producing light elements, and 
though show that they do not resist quantitative analysis.  Looking 
into the reason for this failure, we argue about possible ways to 
alleviate the major problems and propose an alternative scenario based 
on the acceleration of the enriched material filling the interior of 
superbubbles.  We adopt a `conservative picture' for the metallicity, 
in which the Galactic abundances of O and Fe are assumed to be 
proportional.  It should be kept in mind, however, that this picture 
is being put seriously into question by recent observational analysis 
(Israelian et al.  1998; Boesgaard et al.  1998; Garc\'ia-Lopez, these 
proceedings), so that the primary behavior of Be and B \emph{relative 
to their main progenitor, O}, may not be as firmly established as one 
had thought.  In particular, Fields and Olive (1999) have argued that 
the canonical secondary process involving Galactic cosmic rays (GCRs) 
cannot be completely ruled out on the basis of current data (see also 
Fields, these proceedings).  This difficulty could however be regarded 
as another reason to concentrate mainly on the energetics of light 
element production, as the qualitative behavior itself is not yet well 
established.

The key quantitative feature derived from the data, is the 
approximately constant Be/Fe abundance ratio in the halo stars: Be/Fe 
$\simeq 1.6\,10^{-6}$ (Ramaty et al.  1997).  It suggests that each 
ejection of Fe in the ISM by SN explosion must be accompanied, on 
average, by the corresponding amount of spallative Be production, as 
deduced from the above ratio.  A similar requirement can be obtained 
to compare the Be and O yields to the observed values, by noting that 
the spallation processes under study must at least account for the 
least constraining (i.e.  easiest to explain) point in the data giving 
the Be abundance as a function of [O/H].  This statement is 
independent of any assumption about the genuine primary, secondary or 
any other behavior of Be evolution.  What we are asking to the models, 
at this stage, is not to reproduce the whole Galactic evolution of Be, 
but only one point in the data (which is unquestionably a minimal 
requirement).  As we investigate primary processes here, we know that 
the Be/O abundance ratio they produce will in fact be constant.  
Therefore, we obtain the minimal required yield of Be per oxygen 
nucleus ejected in the ISM by merely dividing the abundance of O by 
that of Be in the least constraining star observed.  Using the most 
recent compilation by Fields and Olive (1999), we obtain $Be/O \simeq 
7.5\,10^{-9}$ (or $\sim 6\,10^{47}$ nuclei of Be per solar mass of O).

\section{Light element production in supernova remnants}

In this paper, we investigate the Li, Be and B production associated 
with the explosion of an isolated SN in the ISM. As the SN explodes, a 
large amount of kinetic energy ($E_{\mathrm{SN}}\sim 10^{51}$ ergs) is 
released, causing two shock waves to develop:

\begin{enumerate}

\item a forward shock, containing an energy of about 
$E_{\mathrm{for}}\simeq E_{\mathrm{SN}}\simeq 10^{51}$ ergs, expanding 
outward and sweeping up the circumstellar medium, which is quite 
close to the primordial gas in the very early Galaxy (metallicity $Z 
\simeq 0$), and

\item a reverse (reflected) shock, containing an energy of order 
$E_{\mathrm{rev}}=\theta_{\mathrm{rev}}E_{\mathrm{SN}}\sim 10^{50}$ 
ergs, directed towards the remnant star and sweeping up the SN ejecta, 
rich in freshly synthesized metals.

\end{enumerate}

As is known from both theory and observations, shocks accelerate some 
of the particles flowing through them up to supernuclear energies 
(i.e.  above the nuclear thresholds of order a few MeV/n), and 
distribute these energetic particles (EPs) over an approximately power 
law spectrum with slope $\sim 2$ in momentum.  The efficiency of the 
acceleration process is generally of order $\theta_{acc}\simeq 0.1$, 
which means that about 10~\% of the shock energy is finally imparted
to the EPs.  Once accelerated, the particles diffuse in the 
surrounding medium and interact with the ambient matter, to produce 
light elements by spallation.  The two shocks just mentioned are thus 
at the origin of two distinct processes for Be nucleosynthesis, which 
we now evaluate.

\subsection{Description of process~1}

In process~1, induced by the forward shock of the SN, particles from 
the ISM are accelerated during the whole Sedov-like expansion phase, 
at the end of which the shock becomes radiative and the acceleration 
efficiency quickly drops.  Assuming that the acceleration process is 
not chemically selective, the composition of the EPs has to be that of 
the ISM, i.e.  essentially primordial gas (devoid of metals) in the 
early stages of Galactic chemical evolution.  Concerning the target 
medium, it has to be realized that most of the EPs are actually 
confined within the supernova remnant (SNR) until the end of the 
Sedov-like phase, as they are trapped inside the `diffusion barrier' 
located just downstream of the shock (the confinement is especially 
efficient for the low energy EPs which are the most numerous and the 
most efficient in inducing spallation reactions).  Indeed, the 
acceleration itself is due to the ability of this downstream region 
(hosting strong magnetic turbulence and waves) to diffuse back ionized 
particles so they can pass through the shock front many times.  As a 
consequence, the target material interacting with the EPs should be 
expected to be made of a mixture of the metal-rich SN ejecta and the 
material already swept-up by the shock, i.e.  metal-free ISM.

The main nuclear reactions involved in this Be production process are 
thus direct spallation reactions in which energetic protons (and 
$\alpha$ particles) interact with freshly synthesized C and O nuclei 
ejected by the SN. It is therefore a primary process, as the total Be 
yield associated with each supernova does not depend on the prior 
enrichment of the ISM, but only on the quantity of O (and C) ejected 
by the supernova and the dynamics of the SNR evolution.  The latter is 
particularly relevant to our calculations, for a number of reasons.  
It should be clear, first, that the process only last as long as the 
diffusion barrier is efficient enough to retain the EPs in the 
interior of the SNR, where metal-rich material is encountered.  Now 
this barrier is expected to drop at the end of the Sedov-like phase, 
$t\equiv t_{\mathrm{end}}$, when the shock becomes radiative and the 
magnetic waves dissipate on a short time-scale.  After 
$t_{\mathrm{end}}$, the EPs are free to leave the SNR and diffuse away 
in the whole Galaxy.  Since the latter is essentially devoid of 
metals in the early stages of its evolution, no significant spallative 
nucleosynthesis can be expected after $t_{\mathrm{end}}$, 
which therefore marks the end of our process~1.

The second reason why dynamics is so important in this study is that 
while the EPs are confined within the expanding remnant, they suffer 
adiabatic losses which lower their energy.  Some of the EPs thus slow 
down to energies below the spallation thresholds, which obviously 
causes a decrease of the Be production efficiency.  These energy 
losses need to be taken into account carefully, and make it impossible 
to use standard steady-state models to calculate the total Be yield 
associated with the supernova explosion.

A third feature which makes the use of time dependent models necessary 
is the dilution effect of the target.  Shortly after the explosion, 
the composition of the SNR material is very rich in C and O, so that 
the Be production efficiency (number of nuclei produced per erg in 
EPs) is high.  But this target material gets poorer and poorer as the 
SNR expands and the ejecta are diluted by the swept-up ISM. For a 
given energy in the form of EPs, the Be production rates therefore 
keep decreasing as time flows from the explosion ($t=0$) to 
$t_{\mathrm{end}}$.  Finally, the acceleration of the EPs itself 
cannot be considered as a stationary process.  Indeed, as the shock 
expands, its power decreases as $1/t$, and the rate at which EPs are 
injected inside the SNR follows approximately the same law.

\subsection{Description of process~2}

The second process of light element production which we consider here 
is associated with the reverse shock of the SN. As already mentioned, 
the particles are then accelerated out of the ejecta, and their 
composition is very rich in C and O (much more than the surrounding 
medium).  As a consequence, most of the Be producing reactions will be 
inverse spallation reactions, i.e.  in-flight spallation of energetic 
C and O interacting with H and He nuclei at rest in the ambient 
medium.  This provides a primary process again, as the metallicity of 
the target has no influence on the total Be yield.

Just as for process~1, the EPs are largely confined within the SNR 
until the shock becomes radiative, at the end of the Sedov-like phase.  
Adiabatic losses must therefore be considered during this phase, 
implying that the use of time dependent models is again required.  An 
important difference with process~1 is that the production of light 
elements keeps going on after $t_{\mathrm{end}}$, as energetic C and O 
nuclei can still be spalled into Li, Be and B while interacting in the 
surrounding metal-free ISM.

The last physical ingredient which is needed to calculate to 
spallation rates during the process is the so-called injection 
function, specifying how many EPs are produced by the acceleration 
mechanism per unit of time, and with what energy spectrum.  As for 
process~1, we use the standard shock acceleration spectrum, normalized 
to 10~\% of the shock power.  Acknowledging the
fact that the lifetime of the reverse shock is short compared to the 
other relevant time-scales, we assume that the injection of the EPs 
takes place instantaneously at the sweep-up time, $t_{\mathrm{sw}}$.

More details about processes~1 and~2 and the injection functions will 
be find in Parizot and Drury (1999a,b), together with extensive 
numerical estimates and a detailed discussion of the various 
parameters involved.  A complete description of the time dependent 
model used here can also be found in Parizot (1999).

\section{Numerical results and analysis}

\subsection{Results for process 1}

The Be production rate by process~1 is shown in Fig.~1a as a function 
of time, for different ambient densities.  As already noted, the 
process is efficient only during the Sedov-like phase of the SNR 
expansion, which can be seen to shrink as the ambient density 
increases.  However, the target density is correspondingly higher, 
which implies higher spallation rates as well.  The total, integrated 
yields are shown in Fig.~1b as a function of density for different 
models of SN explosion.  The latter have been taken from Woosley and 
Weaver (1995) (hereafter WW95), and differ in their inputs (initial 
mass and metallicity of the progenitor, explosion energy and velocity 
of the ejecta) and outputs (masses of each element ejected), which are 
relevant to our calculations.  We find that higher densities imply 
larger numbers of Be nuclei synthesized.  However, even in the most 
favorable cases, the numbers obtained are still at least two orders of 
magnitude lower than those implied by the data ($\sim 4\,10^{48}$ 
atoms of Be per supernova; see Ramaty et al.  1997).  The conclusion 
of this quantitative study is that, although process~1 reproduces the 
observed primary behavior of Be (slope 1 in the evolution diagram), it 
cannot be the major source of Be and other light elements in the 
Galaxy.

\begin{figure}
\plottwo{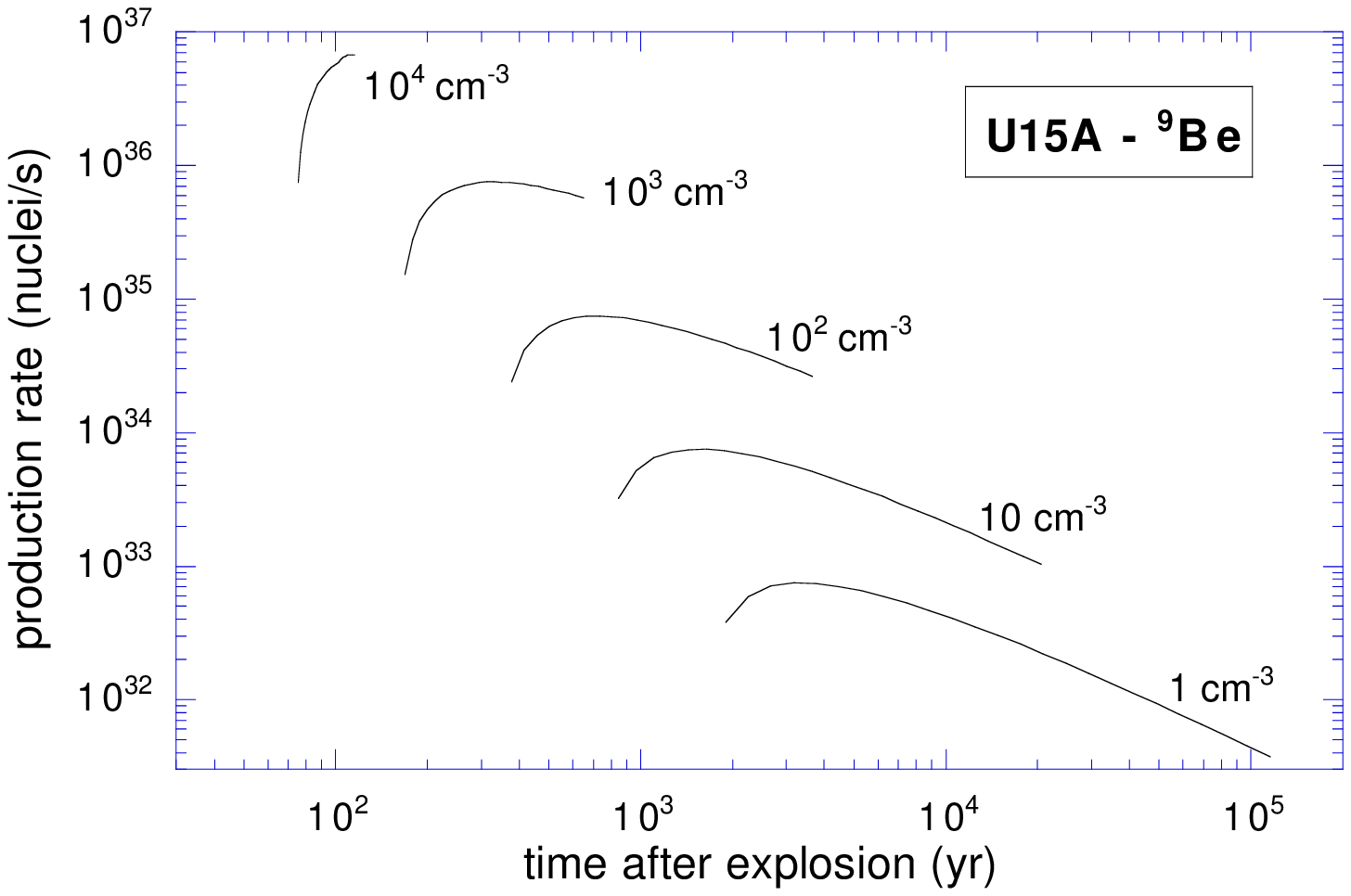}{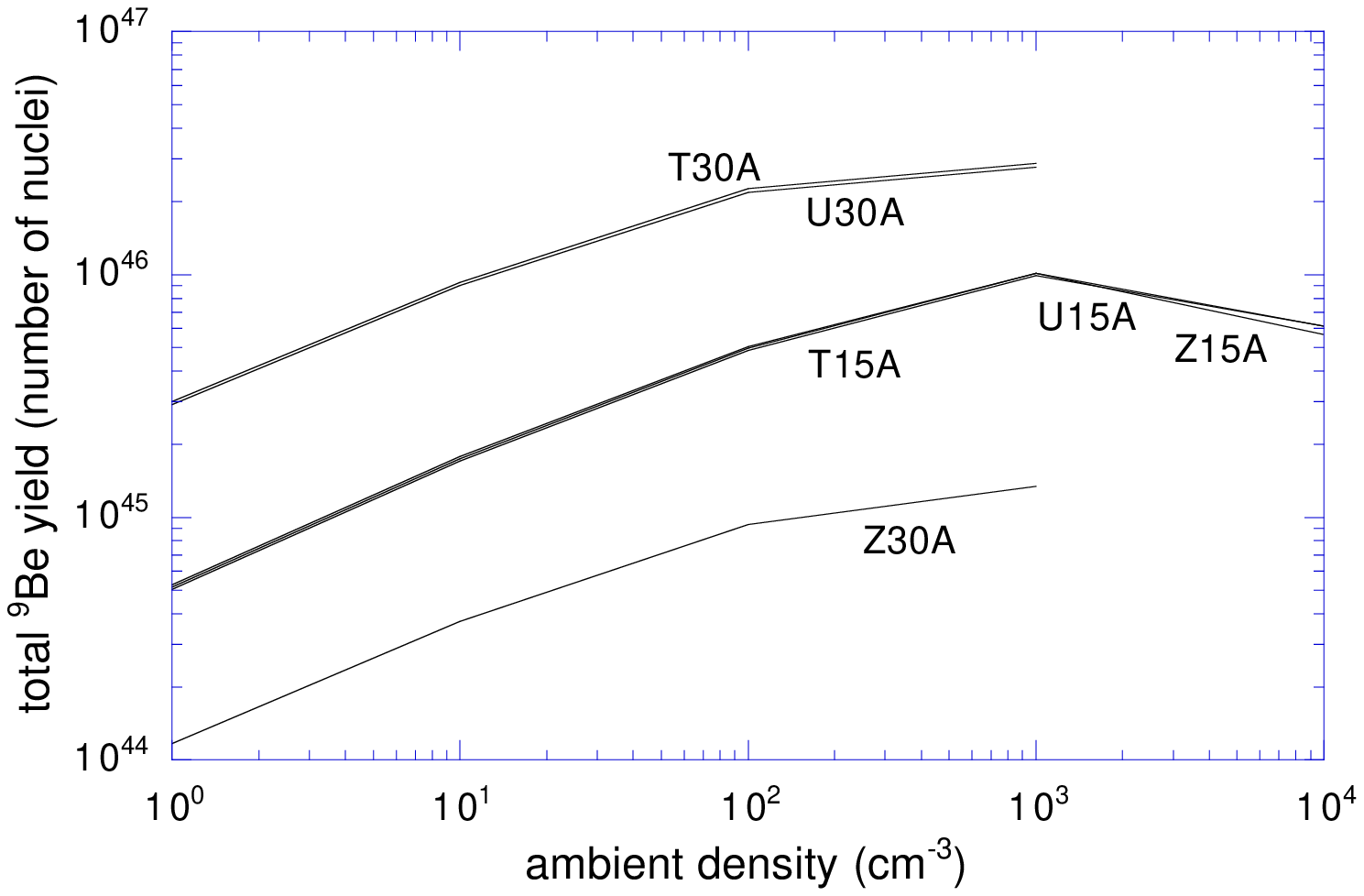} \caption{ (left) Process 1 Be 
production rate as a function of time, for a SN of 15 M$_{\odot}$ 
(model U15A of WW95) and different ambient densities; (right) 
Integrated Be yields obtained by process~1 for the SN explosion model 
U15A (WW95), as a function of the ambient density}
\end{figure}

Analyzing the reason for this failure, we are left with two 
possibilities: either there is not enough energy in the process, or 
the spallation efficiency is too low, that is the C and O-rich ejecta 
are too much diluted by the ambient metal-free gas.  Now this is not a 
small conclusion, as finding a process involving more energy than a 
supernova and metallicities larger than inside a supernova remnant 
seems rather challenging.  it should be noted also that process~1 is 
in any case more efficient than the standard process called Galactic 
cosmic-ray nucleosynthesis (GCRN), in which the forward shock of SNe 
accelerate the ambient ISM, just as in process~1, but the interaction 
with C and O nuclei occurs in the whole Galaxy, where these elements 
are much more diluted than inside a supernova remnant.  As a 
consequence, the failure of process~1 also implies that of standard 
GCRN, at least in the earliest stages of Galactic evolution.

\subsection{Results for process 2}

\begin{figure}
\plottwo{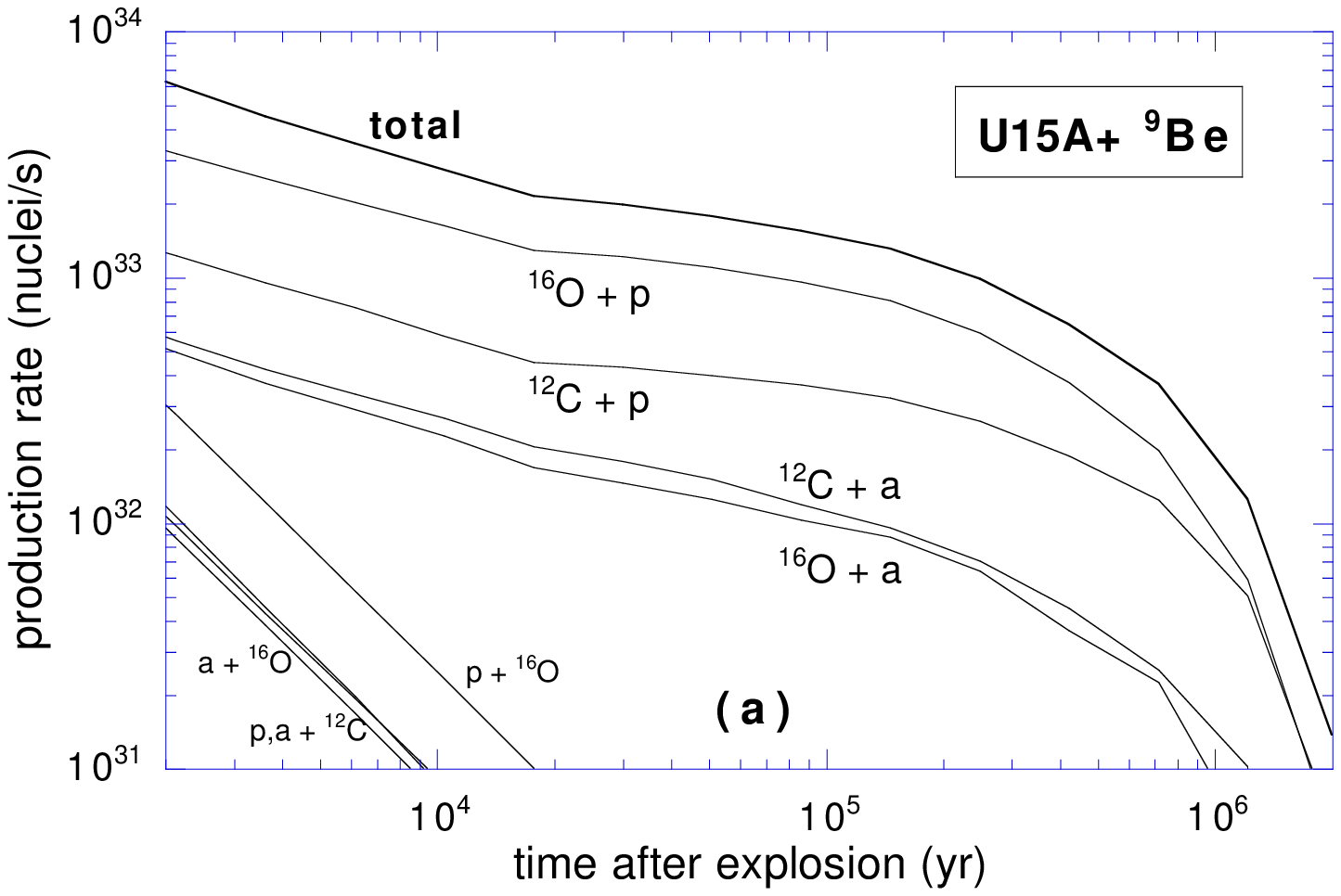}{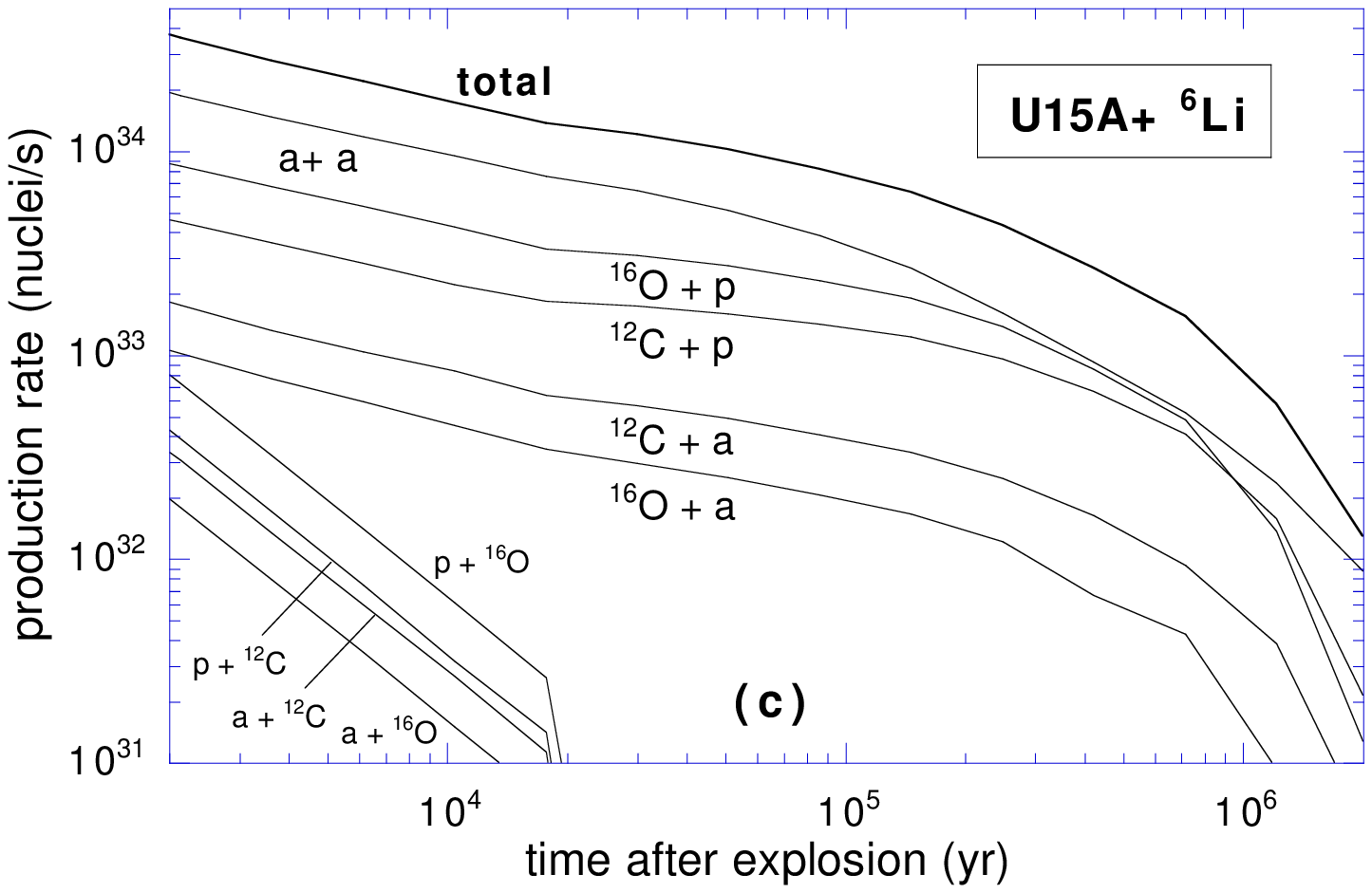} \caption{Detailed $^{9}$Be and 
$^{6}$Li production rates by process~2 as a function of time for a SN 
progenitor of 15 M$_{\odot}$ and initial metallicity $Z = 
10^{-4}Z_{\odot}$ (model U15A of WW95).}
\end{figure}

The time dependent production rates by process~2 are shown in Fig.~2 
for both $^{9}$Be and $^{6}$Li.  Integrated yields, normalized to the 
observationally required values (see Sect.~1) are shown in Fig.~3 as a 
function of the SN progenitor's mass and in Fig.~4 after averaging 
over the initial mass function (IMF), as function of the IMF index ($x 
= 2.35$ for Salpeter IMF).

\begin{figure}
\plottwo{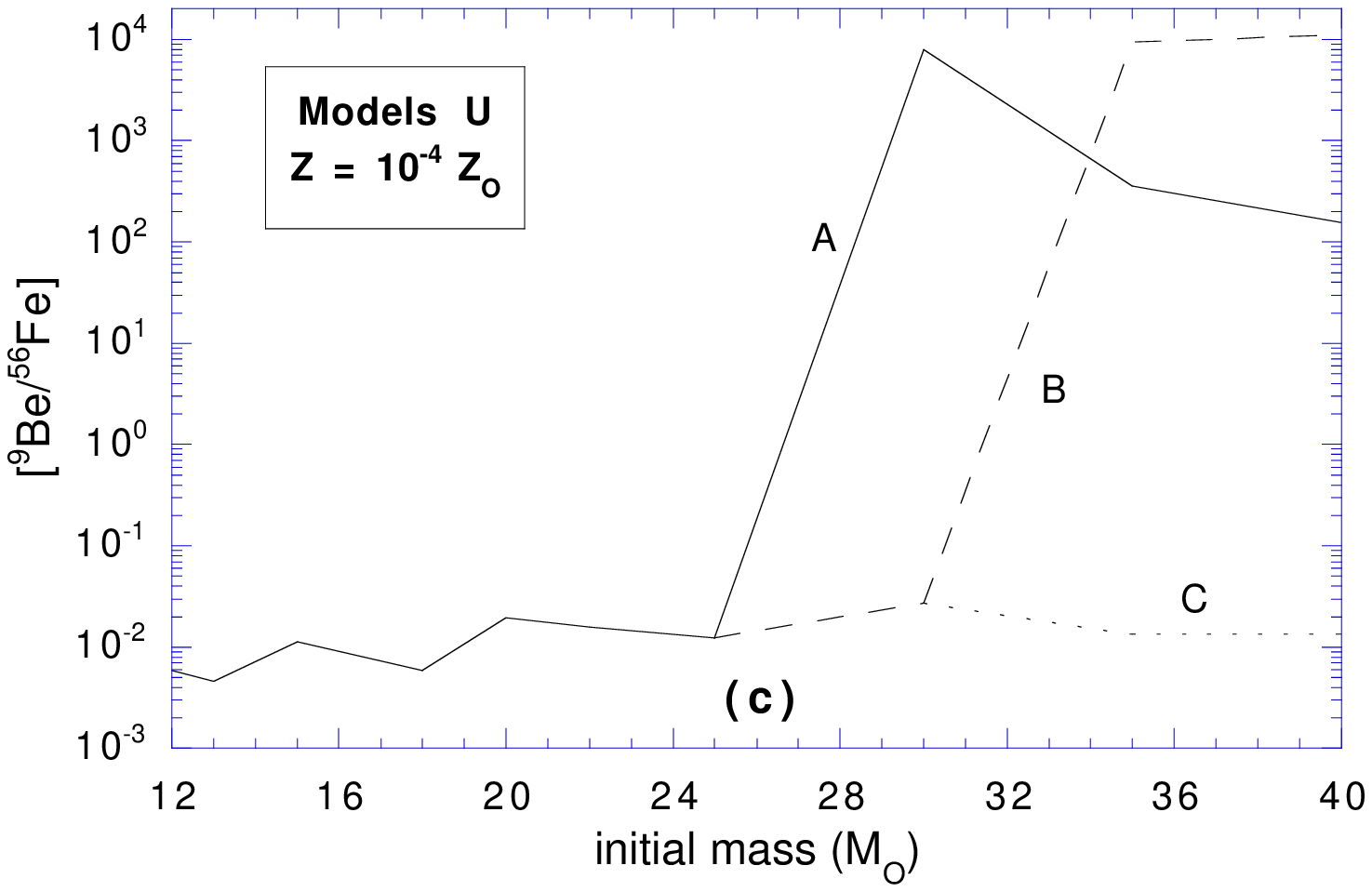}{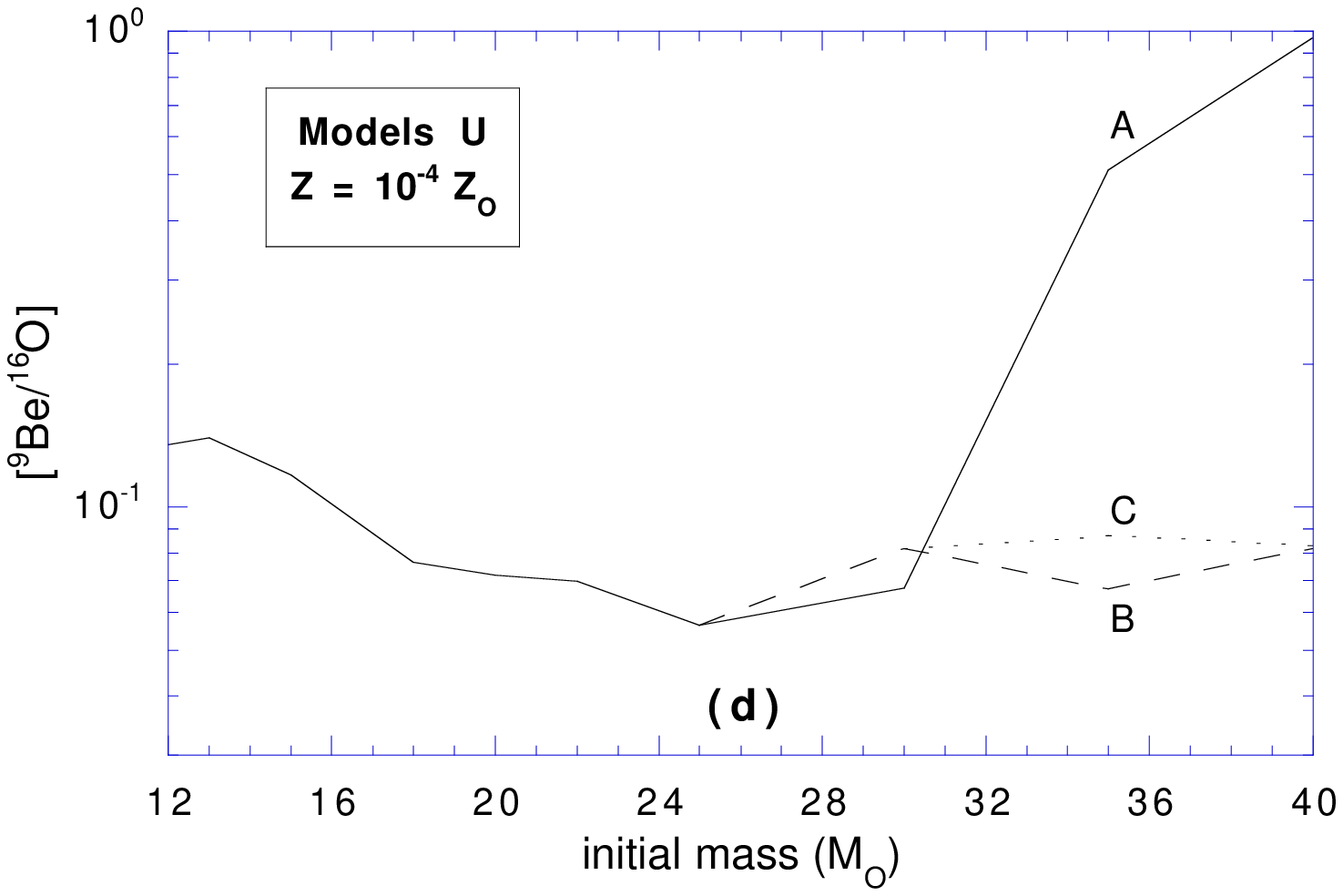} \caption{Process~2 Be/Fe and 
Be/O yield ratios normalized to the values required to explain the 
data (see text), as a function of progenitor's mass, for different SN 
explosion models taken from WW95.}
\end{figure}

It can be seen that process~2 also fails quantitatively, by about two 
orders of magnitude when comparison is made with Fe, and one when it 
is made with O. Note that the latter is the most relevant, as O is the 
direct progenitor of Be and the SN Fe yields may not be well 
understood theoretically.  Normalized Be yields obtained by process~2 
not considering the adiabatic losses are also shown on Fig.~4.  They 
are a factor 3 to 4 higher, which demonstrates the importance of these 
energy losses and the need for time dependent calculations.

\begin{figure}
\plottwo{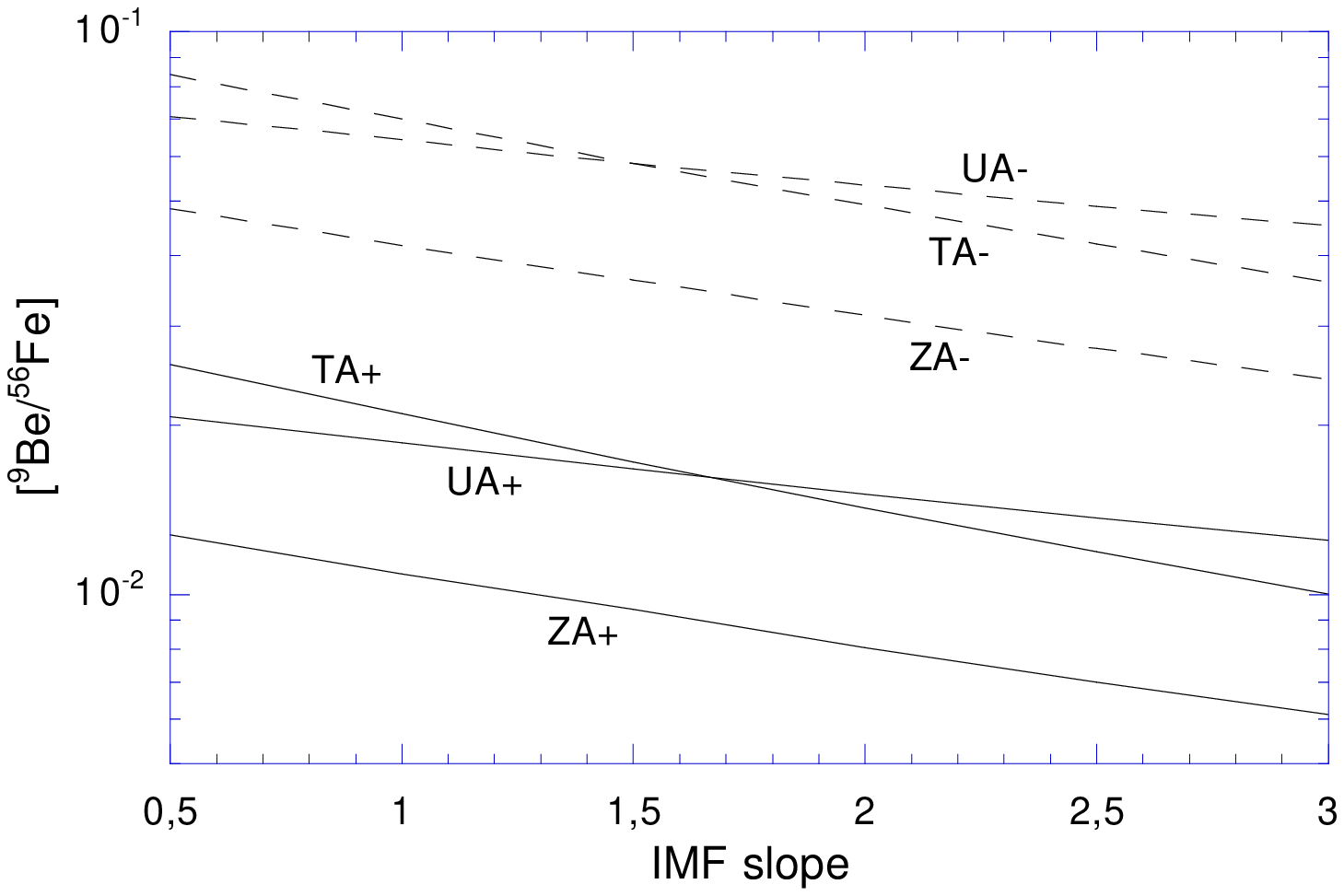}{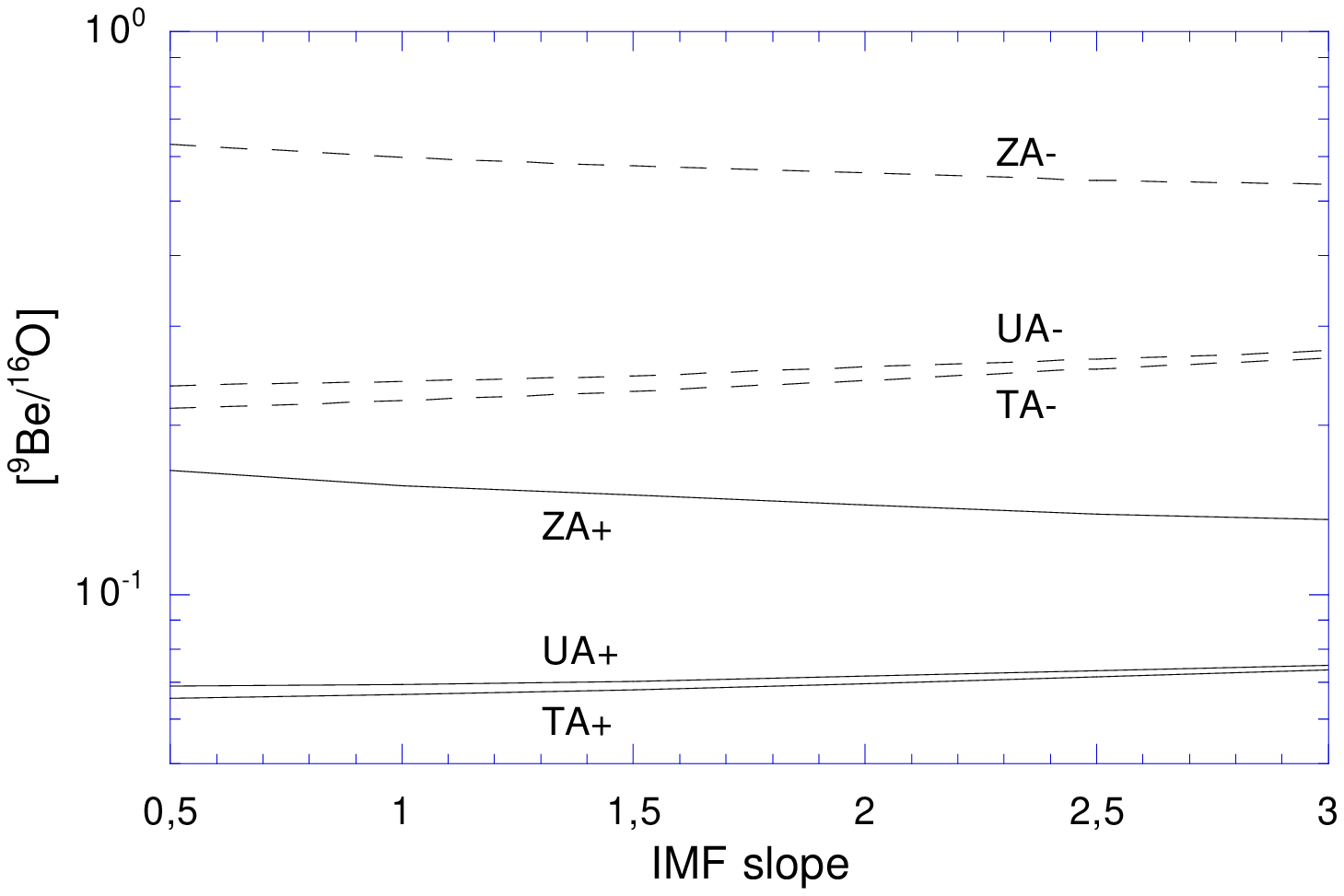} \caption{Be/Fe and Be/O yield 
ratios obtained by process~2 after averaging over the IMF and 
normalizing to the observational values, as a function of the IMF 
logarithmic index.  Dashed lines correspond to the same models with 
the adiabatic losses turned off.}
\end{figure}

\section{Towards a solution of the light element production puzzle}

The results presented in this paper provide important clues towards a 
solution of the Be evolution problem in the early Galaxy.  Our 
process~1 (acceleration of particles from the ISM) fails because the 
target is too poor in C and O, but this cannot be improved.  On the 
other hand process~2 (acceleration of particles from the ejecta) fails 
because of adiabatic losses (factor of 3--4) and because the reverse 
shock is less energetic than the forward shock (factor of $\sim 10$).  
Now both problems may be avoided in a model in which particles are 
accelerated in the interior of superbubbles (SBs), taking advantage of the 
collective effect of SNe in an OB-association, instead of isolated 
SNe.  In such a superbubble model (Parizot et al.  1998, Higdon et al.  
1998), particles are accelerated out of the enriched material ejected 
by earlier massive stars (through winds and SN explosions), just as in 
our process~2, but this is now done by the forward shock, instead of 
the reverse one.  A factor of about 10 in energy could therefore be 
gained.  Moreover, adiabatic losses may be avoided because of the low 
expansion rate of an evolved superbubble.  This would provide an other 
factor of 3, pushing the Be yields at the level of the required 
values, derived from the observations.

Additional work is however needed to work out the details of an 
effective SB model.  The main uncertainties pertain to the composition 
of the EPs and to their acceleration mechanism.  Parizot (1998) and 
Parizot et al.  (1998) argued that the `accelerable material' in a SB 
is made of the averaged wind and SN ejecta of the most massive stars, 
and that the EP spectrum is hard, with a low-energy cut-off (a few 100 
MeV/n), as arises from the SB acceleration models by Bykov and 
Fleishman (1992).  Higdon et al.  (1998) also adopted a SB model and 
justified the previous assumption about composition by geometrical 
arguments, but used the usual shock acceleration mechanism, and thus 
the usual cosmic-ray spectrum.  Now although we all would like to 
accelerated enriched material, we cannot be sure that this is actually 
the case in a SB until the mixing of the stellar ejecta with the 
evaporated ISM off the SB shell has been estimated properly.  As for 
the acceleration process, a key question seems to be: what is the fate 
of a SNR shock in a highly turbulent, tenuous and high temperatured 
medium such as the interior of a SB? If the SNR is essentially 
unaffected, then we should expect standard shock acceleration and a 
typical CR spectrum.  On the other hand, if the energy released by the 
explosion turns into turbulence on a short time-scale, then the 
acceleration mechanism proposed by Bykov and co-workers should be 
adopted, leading to a different energy spectrum, and thus to a 
component of EPs distinct from the ordinary CRs.  This crucial 
questions will be addressed in future works.

\acknowledgments This work was supported by the TMR programme of the 
European Union under contract FMRX-CT98-0168.

\end{document}